\documentclass[conference]{IEEEtran}
\IEEEoverridecommandlockouts
\usepackage{cite}
\def\BibTeX{{\rm B\kern-.05em{\sc i\kern-.025em b}\kern-.08em
    T\kern-.1667em\lower.7ex\hbox{E}\kern-.125emX}}

\ifCLASSINFOpdf
  \usepackage[pdftex]{graphicx}
  \graphicspath{{./graphics/}}
\else
\fi

\usepackage{balance}
\usepackage{subcaption}
\usepackage{amsmath}
\usepackage{hyperref}

\usepackage{listings}
\usepackage{xcolor}

\definecolor{codegreen}{rgb}{0,0.6,0}
\definecolor{codegray}{rgb}{0.5,0.5,0.5}
\definecolor{codepurple}{rgb}{0.58,0,0.82}
\definecolor{backcolour}{rgb}{0.95,0.95,0.95}

\lstdefinestyle{mystyle}{
    backgroundcolor=\color{backcolour},   
    commentstyle=\color{codegreen},
    keywordstyle=\color{magenta},
    numberstyle=\tiny\color{codegray},
    stringstyle=\color{codepurple},
    basicstyle=\ttfamily\footnotesize,
    breakatwhitespace=false,         
    breaklines=true,                 
    captionpos=b,                    
    keepspaces=true,
    escapechar=|,
    frame=single,
    numbers=left,                    
    numbersep=5pt,                  
    showspaces=false,                
    showstringspaces=false,
    showtabs=false,                  
    tabsize=2
}

\begin{document}
%
% paper title
% Titles are generally capitalized except for words such as a, an, and, as,
% at, but, by, for, in, nor, of, on, or, the, to and up, which are usually
% not capitalized unless they are the first or last word of the title.
% Linebreaks \\ can be used within to get better formatting as desired.
% Do not put math or special symbols in the title.
\title{On the Importance and Shortcomings of Code Readability Metrics: A Case Study on Reactive Programming}
%\title{The impact of reactive programming on code complexity and %readability: A Case Study}

% author names and affiliations
% use a multiple column layout for up to three different
% affiliations
\author{\IEEEauthorblockN{Gustaf Holst, Felix Dobslaw}
\IEEEauthorblockA{\textit{Mid Sweden University} \\
\textit{Dept.\ of Computer Sciences}\\
\"Ostersund, Sweden \\
holst.gustaf@gmail.com, felix.dobslaw@miun.se}
}

% \author{\IEEEauthorblockN{Author 1, Author 2, Author 3}
% }

% conference papers do not typically use \thanks and this command
% is locked out in conference mode. If really needed, such as for
% the acknowledgment of grants, issue a \IEEEoverridecommandlockouts
% after \documentclass

% for over three affiliations, or if they all won't fit within the width
% of the page, use this alternative format:
% 
%\author{\IEEEauthorblockN{Michael Shell\IEEEauthorrefmark{1},
%Homer Simpson\IEEEauthorrefmark{2},
%James Kirk\IEEEauthorrefmark{3}, 
%Montgomery Scott\IEEEauthorrefmark{3} and
%Eldon Tyrell\IEEEauthorrefmark{4}}
%\IEEEauthorblockA{\IEEEauthorrefmark{1}School of Electrical and Computer Engineering\\
%Georgia Institute of Technology,
%Atlanta, Georgia 30332--0250\\ Email: see http://www.michaelshell.org/contact.html}
%\IEEEauthorblockA{\IEEEauthorrefmark{2}Twentieth Century Fox, Springfield, USA\\
%Email: homer@thesimpsons.com}
%\IEEEauthorblockA{\IEEEauthorrefmark{3}Starfleet Academy, San Francisco, California 96678-2391\\
%Telephone: (800) 555--1212, Fax: (888) 555--1212}
%\IEEEauthorblockA{\IEEEauthorrefmark{4}Tyrell Inc., 123 Replicant Street, Los Angeles, California 90210--4321}}

% use for special paper notices
%\IEEEspecialpapernotice{(Invited Paper)}

% make the title area
\maketitle

\thispagestyle{plain}
\pagestyle{plain}

% As a general rule, do not put math, special symbols or citations
% in the abstract
\begin{abstract}
Well structured and readable source code is a pre-requisite for maintainable software and successful collaboration among developers. Static analysis enables the automated extraction of code complexity and readability metrics which can be leveraged to highlight potential improvements in code to both attain software of high quality and reinforce good practices for developers as an educational tool. This assumes reliable readability metrics which are not trivial to obtain since code readability is somewhat subjective. Recent research has resulted in increasingly sophisticated models for predicting readability as perceived by humans primarily with a procedural and object oriented focus, while functional and declarative languages and language extensions advance as they often are said to lead to more concise and readable code. In this paper, we investigate whether the existing complexity and readability metrics reflect that wisdom or whether the notion of readability and its constituents requires overhaul in the light of programming language changes. We therefore compare traditional object oriented and reactive programming in terms of code complexity and readability in a case study. Reactive programming is claimed to increase code quality but few studies have substantiated these claims empirically. We refactored an object oriented open source project into a reactive candidate and compare readability with the original using cyclomatic complexity and two state-of-the-art readability metrics. More elaborate investigations are required, but our findings suggest that both cyclomatic complexity and readability decrease significantly at the same time in the reactive candidate, which seems counter-intuitive. We exemplify and substantiate why readability metrics may require adjustment to better suit popular programming styles other than imperative and object-oriented to better match human expectations.

\end{abstract}

% no keywords

% For peer review papers, you can put extra information on the cover
% page as needed:
% \ifCLASSOPTIONpeerreview
% \begin{center} \bfseries EDICS Category: 3-BBND \end{center}
% \fi
%
% For peerreview papers, this IEEEtran command inserts a page break and
% creates the second title. It will be ignored for other modes.
\IEEEpeerreviewmaketitle

\section{Introduction}
\label{introduction}
Code readability and complexity are both considered to be important aspects in software development as they directly link to code maintainability \cite{buse_learning_2010, posnett_simpler_2011, alawad_empirical_2019}. Code maintainability is one of the critical code quality characteristics defined by the standard model for assessing software quality in ISO/IEC 25010:2011 \cite{noauthor_iso_nodate}. With the appearance of every new programming style or paradigm, software developers and researchers are interested in finding empirical evidence for its actual attributes, and benefits in particular. Chidamber and Kemerer \cite{chidamber_metrics_1994} as well as Abreu and Carapuca \cite{redol_object-oriented_nodate} have contributed by establishing metrics for object-oriented programming (OOP) incorporating characteristics such as coupling and cohesion. As pointed out in \cite{redol_object-oriented_nodate}, a new programming paradigm will only gain widespread practical acceptance when its management aspects are carefully addressed. Developers can leverage that knowledge to make informed technological choices, while researchers are supported in their design decisions for the creation of the programming languages of the future.

One of the claims of the reactive programming (RP) paradigm is that it promotes readability. RP extensions to existing high-level languages, such as Reactive Extensions\footnote{\url{http://reactivex.io/}. The Java version, \textit{RxJava}, has as of now over 40k stars on GitHub which places it in the top 10 of Java projects.}, are increasingly used due to their consistent use across language barriers and simplified multi-threading, amongst others. The work of Salvaneschi et al. indicates that RP has an advantage over the traditional OOP approach when it comes to program comprehension \cite{salvaneschi_positive_2017}. Their study used human subjects rating two versions, one imperative and one reactive, of the same code excerpts. High adaptation therefore highlights the need to understand whether existing static analysis readability metrics are capable of capturing high readability. Or put another way, does RP really improve readability and if so, can human-noticable differences be detected with current readability metrics?

Buse and Weimer in \cite{buse_learning_2010} investigated the relationship between code readability and software quality. To this end, they first collected data using 120 human subjects rating short code snippets and then developed a mathematical model to estimate the readability of code, which we here refer to as the B\&W model. The B\&W model has been expanded upon by \cite{posnett_simpler_2011, dorn_general_2012, scalabrino_improving_2016}, and Scalabrino et al. in \cite{scalabrino_comprehensive_2018} propose a combined model, the Scalabrino Comprehensive (SC) model, that they show to be more accurate at assessing code readability as perceived by humans. Both these models are created using machine learning techniques and trained on imperative code excerpts. However, the features used are generic with, for example, low level features such as number of periods and parentheses or visual and textual features that affect readability and are in no way specifically designed for either imperative or object oriented programming. The assessment of code-readability has nothing to do with the choice of programming language and everything to do with the comprehension of programs in general. If a metric is designed biased towards a certain programming language/paradigm, it will empirically fail to capture high readability, which is what we argue here is undesirable.

While \cite{salvaneschi_positive_2017} studied comprehension effects of RP involving human subjects, we seek to know to what extent changes in readability between OOP and RP can be detected, using common static code analysis. To the best of our knowledge, the work in \cite{salvaneschi_positive_2017} is the only attempt at studying the difference in comprehensibility between OO and RP and the authors call for more research in the field. It is important here to point out the distinction we make between \emph{comprehensibility} and \emph{readability}. In lack of a commonly accepted definition, we consider software comprehensibility to be a higher-level property of a software system than code readability. Comprehensibility may relate to how the structural relationship in software can be understood while we consider readability to be assessed on a lower level, a level at which code is processed by the cognitive apparatus of a human, i.e. methods or functions. Both interlink and correlate, but are not the same thing.

Our study is focused on comparing two versions of the same code excerpts, namely the original imperative version and a declarative (functional) version leveraging abstractions of a reactive framework. The two tools we will use for comparing readability, the implementations of B\&W and SC, has limitations that affect our methodology. The B\&W tool seems limited excerpts of around 7 lines, longer excerpts are likely to yield a readability score of 0.0. The SC tool on the other hand can provide a readability score on class level but it is then calculated as a mean of the readability scores for the individual methods. Also, SC requires excerpts to be well-formed (compilable). We will refactor and compare at method-level allowing us to, one, use the mentioned tools, and two, compare excerpts performing the same tasks.  A true reactive system would be designed differently than our refactored version but we are interested in the \emph{readability} of the reactive code and more specifically if its declarative style, perhaps unjustly, penalizes readability.

The paper is structured as follows. The problem and research questions are detailed in Section \ref{problem}. In section \ref{limitations} we acknowledge and discuss the limitations of the study. The Background section in \ref{background} prepares the reader with a basic understanding of reactive programming, as well as readability and code complexity measures, for which we summarized the related work in Section \ref{relatedwork}. In Section \ref{method}, we detail the conduct of the case study, for which we present the results in Section \ref{results}. The research questions are answered in Section \ref{discussion}, Discussion. A summary of the work with concluding remarks can be found in Section \ref{conclusions}, Conclusions.

\section{Problem Statement}
\label{problem}
The importance of code readability and code complexity reflects from its strong influence on other quality metrics, and is considered a key element in software development \cite{mduniversity_code_2017}. In particular, increased readability eases maintenance, which is often considered the most time consuming activity of a software's life-cycle \cite{alawad_empirical_2019}. Salvaneschi et al. show empirically how RP may improve comprehensibility but stress the need for further research \cite{salvaneschi_positive_2017}.
To the best of our knowledge, there is still little to no research concerning the effect of readability and complexity when comparing RP to OOP. Comments from the human subjects in \cite{salvaneschi_positive_2017} suggest that RP improves comprehension due to reduced boilerplate code and increased readability.

The goal of this study is to  to better understand whether current state-of-the-art metrics are applicable to RP or tailored RP metrics need to be developed. To this end we use a complementary research method to \cite{salvaneschi_positive_2017}, where instead of using human subjects to provide a gold standard we simply rely on the existing state-of-the-art models for measuring readability to tell us whether a method is readable or not. We then search for evidence that supports the hypothesis that RP is more readable or more comprehensible compared to traditional OO source-code, and discuss whether the results justify its marketing as such or not. We chose RP for reference as it has greatly grown in popularity and combines both features of declarative and functional programming.

\subsection*{Research Questions}
RP is claimed to improve code quality in various aspects. In this work, we investigate how RP affects the readability of source code when compared to OOP using existing readability metrics, leading to the first research question:

\emph{RQ1. Using state-of-the-art readability metrics, to what degree can we measure a difference in readability between OOP and RP?}

The second research question examines if the same metrics are applicable for both RP and OOP, or whether the metrics are biased or need further examination. Furthermore, code complexity has shown to be closely related to, and traditionally often used as, a substitute index for code readability. We are interested in what characteristics, if any, OOP or RP offer respectively rendering them advantageous over the other when it comes to readability:

\emph{RQ2. What features, if any, in current code readability metrics must be modified or excluded in order to make them applicable to RP?}

With RP being a fairly new paradigm, no metrics have yet been developed aimed towards it. Thus, metrics used for measuring source code readability might not be well suited for RP. If so, it may be necessary to change weighting of features or exclude them in order to more accurately assess both readability and code complexity of RP code bases.

\emph{RQ3. How does RP, with its declarative nature, affect code complexity compared  to OOP?}

When measuring source code readability and complexity, we hypothesize that RP yields higher readability and lower complexity scores. However, some features of RP might affect the score negatively. For instance, one characteristic of RP (as well as functional programming in general) is its use of so called \textit{chaining} where several function calls are chained resulting in very long expressions at times. While this feature might influence the score of some metrics negatively its declarative style is generally considered a way to increase readability and flexibility.

\section{Limitations}
\label{limitations}
In this case study, we refactor an open source project, Kryonet, written in a traditional imperative style, to a version using reactive pipelines. The chosen project was based on the criteria mentioned in Section \ref{method}, Research Methodology. The focus during the translation process has been to map the original code as closely as possible into a reactive version in order to keep the two versions comparable. Furthermore, efforts were made to keep the original structure of the code. In other words, no functional decomposition was performed and the interface of the refactored library was kept intact. Thus, the refactor cannot be said to constitute a paradigm shift from imperative to reactive as far as design goes and the true benefits of RP might not be fully utilized in the same way as in a project developed with reactivity in mind already during the design phase. However, with our definition of readability, and its relation to comprehensibility as discussed in Section \ref{introduction}, we argue that we \textbf{can} draw conclusions on method-level readability from the comparisons. Also, a method-level refactor might be considered the first step in a paradigm shift from imperative to reactive, though, for reasons given in \ref{method} the subsequent steps were not carried out in this study. The RP library used in this study is RxJava due to its popularity, thus limiting us to the features of its API.

Kryonet goes under the Revised BSD License (BSD 3) which permits unlimited redistribution for any purpose as long as its copyright notices and the license's disclaimers of warranty are maintained. 

\section{Background}
\label{background}

\subsection{Reactive Programming}
Reactive programming is a declarative programming paradigm that has been proposed to aid programmers develop event-based applications. A major issue in such applications is the inherent unpredictability in the timing of external events. Traditionally this has been solved by using so called callbacks, often leading to what is known as \textit{callback hell} as projects grow \cite{bainomugisha_survey_2013}. RP programming languages and libraries provide abstractions over time management to release the burden of the programmer. Any change of states within the reactive circuit is automatically propagated to all dependencies which keep all the affected variables and objects up-to-date \cite{bainomugisha_survey_2013}. A well-known example of the reactive principle is a common spreadsheet where a change in one cell automatically causes updates in other cells that are linked to it. RP builds around events and streams that are chained together using various sets of abstractions to perform operations and computations. When operating in an asynchronous environment, RP is seen as the better alternative compared to traditional object oriented use with thread pools due to its declarative and concise syntax for composing streams. Multiple languages support reactive programming and numerous library extensions have been developed such as Reactor, Akka and Reactive Extensions (Rx) \cite{banken_debugging_2018}.

\subsection{RxJava}
ReactiveX (Reactive Extensions) is a library for reactive programming with implementations in several languages, RxJava being the Java implementation. It is built around the observer design pattern and provides operators with which the programmer can compose reactive streams in a declarative style. Precarious elements of asynchronous programming such as thread-management, thread-safety and synchronization are handled by the library through abstractions in the form of designated operators.

RxJava promotes a functional programming style, as opposed to an imperative one (using callbacks). The main difference from other functional APIs is that where for example the Java Streams API is synchronous and pull based RxJava is asynchronous and push based, meaning data is pushed to any observers whenever available. The core type in RxJava is aptly named \emph{Observable}. An observable stream can emit three kinds of events, namely, a successful emission of data, an error or a completion event signalling that no more emissions are to be expected.

\subsection{Readability}
The term readability stands out among the general software metrics as it refers to a non-physical characteristic, i.e. the subjective human judgement of how easy a code is to read \cite{tashtoush_impact_2013, posnett_simpler_2011}. Plenty of metrics for measuring the readability of general text exist and the Flesch-Kincaid metric from 1948 is only one but possibly the oldest example \cite{flesch_new_1948}. Until recently, with the lack of scientifically substantiated readability metrics specifically aimed at code, various complexity metrics have been used to assess software maintainability \cite{mccabe_complexity_1976, halstead_elements_1977}. Complexity is closely coupled to the problem at hand which is problematic since a more complex problem will require more complex code. Also, code size is commonly used as a metric for complexity but it cannot be taken for granted that more lines of code influence readability negatively at all times. As Posnett et al. point out in \cite{posnett_simpler_2011}: ''\textit{if two functions are equally readable then why should their concatenation be less readable?}``.

Already in 1976, McCabe introduced his influential work on cyclomatic complexity (CYC) as a measurement of code complexity \cite{mccabe_complexity_1976}. It uses a simple formula which in essence counts the number of possible paths through the code:

\begin{align}
    CYC = E - N + 2P,
\end{align}

with
\begin{description}
    \item[E] being the number of edges in the flow graph,
    \item[N] being the number of nodes in the flow graph, and
    \item[P] being the number of nodes that have exit points.
\end{description}

CYC has widely been criticized \cite{shepperd_critique_1988, pizka_how_2007, noauthor_cyclomatic_nodate}. Shepperd \cite{shepperd_critique_1988} argues that it is ''\textit{based upon poor theoretical foundations and an inadequate model of software development}``. 
% It is also important to note that it was developed for the Fortran environment and does not take into account lambdas or other later inventions.
However, as of today, it remains a much cited and utilized index for code complexity.

According to \cite{pantiuchina_improving_2018},
source code complexity is defined by ''\textit{the degree to which a system’s design or code is difficult to understand because of numerous components or relationships among components}``. Code complexity is said to influence a large proportion of other quality factors, e.g. extensibility, maintainability, readability, consistency and understandability. Complexity is therefore considered a key characteristic when speaking of code quality. Multiple metrics concerning code complexity have been proposed over the years, with cyclomatic complexity, lines of code, and Halstead Complexity being the most common ones.

\section{Related work}
\label{relatedwork}
\subsection{The positive effects on comprehensibility of reactive programming}

Salvaneschi et al. performed the first empirical study addressing the claim that RP ''\textit{makes a wide class of otherwise cumbersome applications more comprehensible}`` \cite{salvaneschi_positive_2017}. They set up an experiment with ten, what they call \textit{comprehension tasks}; short programs written in Scala using the reactive library REScala. Human subjects were first presented with these tasks and later asked to fill out a questionnaire aimed at determining the level of comprehension acquired for each task. A task consisted of two versions of the same reactive application, a traditional OOP version using the Observer pattern and an RP one using reactive abstractions such as signals and events. Three categories of applications were  developed:
\begin{itemize}
\item \emph{Synthetic applications} where functional dependencies are defined between variables and changes to those variables are propagated as updates occur,
\item \emph{Graphical animations} in which shapes are displayed and moved around on a canvas in certain patterns, and
\item \emph{Interactive applications} requiring the user to perform common GUI related actions such as clicking a button or inserting text into a form.
\end{itemize}
For the conduct of the study, a web application was developed which was used to present the subjects with the comprehension tasks and, importantly, the time needed for each task was recorded.
The selected subjects were all computer science students with similar backgrounds. They had been taught Scala, Java programming, the observer pattern, as well as the basics of RP.
From their results, the authors concluded that RP has an advantage over OOP and the observer pattern when it comes to comprehensibility. They mainly contributed this difference to two characteristics: reduced boilerplate code and better readability.

\subsection{Developing code readability metrics using human subjects} %alt "Research on code readability..."

In 2010 Buse and Weimer \cite{buse_learning_2010} presented the first readability metric for source code. They collected data by letting 120 human subjects rate the readability of 100 short code fragments. The resulting data set was used to create a predictive model for readability based on 25 code characteristics and their results indicate that the average number of identifiers, line length (avg and max), avg use of brackets, parentheses and punctuation, have the greatest negative impact on readability. Finally, they tested for correlation between a snippets readability score calculated by their model and the number of bugs found by a static analysis tool, and found that their readability metric correlated negatively to the number of defects found, i.e. the lower the score the more likely a defect is found. They concluded that readability was not closely related to the traditional notion of complexity while their notion of readability was ''\textit{not orthogonal to complexity, it is in large part modeling a distinct phenomena}``. Posnett et al. \cite{posnett_simpler_2011} suggested a simpler formula using only three features, lines of code, entropy, and Halstead’s Volume metric that they found corresponds more accurately to the data collected in \cite{buse_learning_2010}.
Dorn \cite{dorn_general_2012} later pointed out that the work in \cite{buse_learning_2010} only used very small code snippets (avg. 7 lines) and that it was not clear whether their findings hold true for longer code excerpts. Dorn concluded that visual elements are missing in the B\&W model and proposed a readability model with added features such as structural patterns, syntax highlighting and operator alignment. He also incorporates the simple linguistic feature of counting the number of English dictionary words used in identifiers. The model proposed by Dorn \cite{dorn_general_2012} was shown to outperform the basic B\&W model in terms of code readability assessment.

Scalabrino et al. in \cite{scalabrino_comprehensive_2018} assess that the existing readability metrics (B\&W, Posnett, Dorn) rely mostly on structural features, suggesting the exploration of linguistic aspects. Their proposed model considers textual features such as identifier consistency, comment readability, the number of \textit{meanings} and textual coherence. The authors proposed a model combining all of the individual models above which they showed to outperform each.

\subsection{Deriving a readability score}
The readability metrics from \cite{buse_learning_2010} and \cite{scalabrino_comprehensive_2018} both rely on logistic regression models to calculate probabilistic readability scores. The data collection was done in a similar fashion in both studies, described as follows. Code snippets were rated by humans on a 5-point scale where 1 denotes lowest, and 5 highest readability. Binary readability classifiers were build based on the obtained mean readability score of snippets as cut-off points ($\geq{3.14}$ in the case of B\&W and $\geq{3.6}$ for SC), thus leading to a binary separation of all code snippets into readable or non-readable. For a given data-point, i.e. code snippet, the model then returns how readable it is on a $0$-$1$ normalized scale.

\subsection{Detecting changes using static analysis}
Pantiuchina et al. \cite{pantiuchina_improving_2018} investigate to what degree state-of-the-art static analysis tools capture source code improvements. The metrics they considered were cohesion, coupling, complexity and readability. They applied repository mining to extract commits from GitHub where the committer explicitly states the purpose of the change to address one of these characteristics, and checked through static analysis whether the change was noticeable. They concluded that neither CYC, nor B\&W or SC readability, could adequately detect any improvements. These results has been confirmed in a similar study by Fakhoury et al. \cite{fakhoury_improving_2019} in which they even observed cases where readability as measured by state-of-the-art readability metrics decreased after a readability commit which indicates that there might be a discrepancy between readability as defined by the metrics and as defined by programmers.

Scalabrino et al. investigated in \cite{scalabrino_automatically_2019} the possibilities of measuring code comprehensibility using static analysis. A number of proxies for understandability were established including reader's own perception of comprehension, actual comprehension and time time needed for perceived comprehension. They tried to correlate 121 code-related, documentation-related and developer-related metrics with these proxies for understandability but without success, i.e. not even readability was found to correlate directly with understandability. Among the evaluated metrics were CYC and SC as well as individual features of B\&W.

\subsection{Correlating readability with cognitive load and design quality}

The possibility of using brain imaging techniques and eye tracking devices to measure the cognitive load of readers of code has been studied by Fakhoury et al. \cite{fakhoury_effect_2018}. Human subjects were asked to fix bugs in code excerpts altered in such a way that either lexical inconsistencies, poor structural characteristics and code features that negatively affects readability were introduced. Table \ref{tab:fakhoury_features} lists the features used to artificially manipulate the readability of code excerpts.

\begin{table}[ht]
\centering
\begin{tabular}{lccc}
    \textbf{Feature} & \textbf{Correlation} \\
    \hline
     CYC & $---$  \\
     number of arguments & $---$ \\
     LOC & $---$ \\
     max nesting depth & $--$ \\
     number of loops & $-$ \\
     number of expressions & $-$ \\
     number of statements & $-$ \\
     variable declarations & $-$ \\
     number of comments & $++$ \\
     number of comment lines & $++$ \\
     number of spaces & $+$ \\
     \hline
\end{tabular}
\caption{Metrics used by Fakhoury et al. to alter the readability of code snippets and their correlation levels and directions. }
\label{tab:fakhoury_features}
\end{table}

The results indicate that lexical inconsistencies (linguistic anti-patterns) increase the cognitive load but its is not evident that poor structure (non standard formatting) or poor readability has the same effect. However, the combination of lexical inconsistencies and/or poor structure/readability had a significant effect on comprehension.

Mannan et al. investigated readability in OSS projects. The study presented in \cite{mannan_towards_2018} focus on how readability changes during the lifetime of a project and also how it impacts design quality. They found that readability, measured using the Posnett model, does not correlate with design quality, measured by amount of code smell using static analysis.

\subsection{Impact of lambda expressions on readability and complexity}
Lucas et al. \cite{mendoncca2020understanding} conducted an extensive study on lambda expressions and the impact they have on readability and complexity. Both a qualitative study using human subjects rating, and commenting on, the readability of code excerpts and a quantitative study measuring readability using two state-of-the-art readability models were carried out. The readability models used were B\&W and the simplified model proposed by Posnett and the results (see Figure \ref{tab:lucas_results} for summary) indicate that neither model is able to capture readability as perceived by humans on code excerpts containing lambda expressions.

\begin{table}[ht]
\centering
\begin{tabular}{l|c|c|c}
    \textbf{Evaluator} & \textbf{Increased} & \textbf{Decreased} & \textbf{Unchanged} \\
    \hline
     B\&W & 23 & 32 & 3 \\
     Posnett et al. & 11 & 43 & 4 \\
     Humans subjects & \textbf{51} & 3 & 4 \\
     \hline
\end{tabular}
\caption{Number of code excerpts that had an increase, decrease in readability or were left unchanged after replacing anonymous inner classes with lambda expressions by refactoring tools.}
\label{tab:lucas_results}
\end{table}

\subsection{Contributions of this paper}
To the best of our knowledge, the work of Salvaneschi et al. in \cite{salvaneschi_positive_2017} is the only study to date that addresses the positive effects of RP. Their study involves human subjects to determine whether a program written using RP principles yields more comprehensible code than a version of the same program written using OOP principles. We intend to tie together the work of Salvaneschi et al. and that of B\&W and SC to assess differences in readability. For that sake, an existing open source project was refactored using Reactive Extensions (Rx). Besides using the readability metrics, we further obtain source code complexity readings through CYC to better understand whether it may be suitable for RP.

The work of Lucas et al. \cite{mendoncca2020understanding} is focusing on readability of code excerpts containing lambda expressions. Reactive programming makes heavy use of lambda expressions which is why understanding their impact on readability and comprehensibility is important in this scope. Furthermore, while Lucas et al. measured readability with B\&W and the Posnett model, our study leverages SC and B\&W, with the Posnett model being incorporated into SC.

\section{Research methodology}
\label{method}

In this section, we first provide a brief outline of the applied methodology and then go on to detail each step of the study.

The reactive library of choice is RxJava from Reactive Extensions, i.e., specifically its Java implementation in version 3.0.2. 
RxJava is among the most popular Java projects on GitHub at the time of this writing, and the most popular of Reactive Extensions considering GitHub stars.

Firstly, a suitable open source project for the study had to be selected. Once a project was shortlisted which met the criteria explained in the upcoming section, it was forked and refactored according to the RP paradigm, also further explained below. 

The original code was refactored on a per-method basis and once a complete reactive version passed all tests in the test suite, both candidates got processed through static analysis gathering CYC complexity and readability according to the B\&W and SC models.
% For this purpose, an application was developed, using JavaParser\footnote{https://javaparser.org/}, which gathered CYC complexity and readability  according to the B\&W and SC models.
The results were extracted and compiled in combination with those features of highest predictive power according to the B\&W model for analysis, namely the \textit{average number of identifiers}, the \textit{average line length}, the \textit{average number of parentheses and brackets}, as well as the \textit{maximum line-length} and \textit{average stops count}. The refactored source code and the measurements can be viewed in the replication package \footnote{https://doi.org/10.5281/zenodo.4277872}.

\subsection{Project choice motivation}
The criteria that guided the search for the case project were the following:
\begin{itemize}
    \item Java project (to enable the use of RxJava)
    \item Associated test suite (for ensuring valid refactor)
    \item Headless (to facilitate testing)
\end{itemize}
RP is said to facilitate the construction of event-based applications. With that in mind we decided to search for a network library in which incoming and outgoing events are handled. The choice of project fell on Kryonet (see Table \ref{tab:project_size}), a TCP\slash UDP client\slash server library written in Java 7. 
Functional features such as Lambdas got introduced in Java 8, which is why Kryonet did not contain any of the functional elements, which rendered it a suitable candidate. Kryonet further has a test suite to ensure valid refactoring and has about 1.5k stars on GitHub at the time of this writing.

\begin{table}[ht]
\centering
\begin{tabular}{l|c|c|c}
    & \textbf{num classes} & \textbf{num methods} & \textbf{loc} \\
    \hline
     Kryonet & $ 54 $ & $ 283 $ & $ 3664 $ \\
     \hline
\end{tabular}
\caption{Size metrics for the original system.}
\label{tab:project_size}
\end{table}

\subsection{Refactor strategy}

Refactoring was conducted manually, i.e., not using any automated refactoring tool, since no such tool was found to be adequate for the purposes of this study. We chose to refactor the original code method per method, leaving the public interface of the library intact. An alternative to this approach was a complete overhaul of the code using RxObservables through the API. Three main reasons justify our choice to refactor per method instead. Firstly, it allowed the use of the test suite as is. Changing the public interface requires the creation of new tests which implies a risk of introducing faults. Secondly, the code sections that are compared are those accomplishing the same task which substantially simplifies the comparison. Lastly, both B\&W and SC readability models were trained on smaller code snippets (average line length of 7), and likely do not generalize well to larger constructs, such as classes.

A number of methods were excluded from refactoring because of one or more of the following reasons.
\begin{itemize}
\item Number of  Lines $>$ 50 (cutoff used in \cite{scalabrino_comprehensive_2018}).
\item No iteration over a Collection present.
\item No error handling present.  
\item Methods containing a single statement.
\end{itemize}
The intention during refactoring was to place all statements in a method in a reactive stream, i.e., into a chain of operators. For operations that need to be performed once at the beginning or at the end of the code block, two side-effect RxJava operators were utilized, \textit{doOnSubscribe} and \textit{doFinally}. Other statements performed in loops were placed in \textit{doOnNext} operators with each statement being placed in a separate operator of the chain for reasons of flexibility. For example, multithreading may be introduced anywhere in the chain by simply adding an extra operator at the desired point. A single long chain further allows us to get rid of try/catch blocks altogether in favor of RxJava's error handling channel for error delegation downstream. In a pure migration project, many code segments would have been extracted and placed in separate methods (functional decomposition) in order to bring out the declarative qualities of RxJava. Such refactoring would clearly have benefited the readability of the reactive stream but made for a misaligned comparison to the original code. Overall, many original code segments simply had to be placed in \textit{doOnNext} operators. For example, the original code base contains many log-level guarded logging statements. We placed these statements in \textit{doOnNext} operators in a way that is arguably not considered preferred lambda syntax \cite{vanags_perfect_2018}, i.e., using the shortest possible definition for the sake of readability.

The convention used for naming lambda parameters was to use single characters. This choice was inspired by Bob Martin in \cite{martin_clean_2008} who argues that the length and descriptiveness of a variable should be in correlation to the size of its scope.

The refactoring process resulted in 42 refactored methods from five different classes and the success of the refactoring was confirmed by running and passing the test suite.

\subsection{Data Collection}
We performed our study as a case study in which we manually refactored an existing OOP code base to RP. One alternative to this approach would have been to find a large number of open source projects of two categories, purely OOP and purely RP, and measure and compare general readability between the two categories. However, OOP projects are plentiful on GitHub, but finding pure RP projects is difficult. Also, pinpointing the actual cause of any differences in readability is hard as multiple factors, such as complexity of the problem being solved or general coding style (naming conventions etc.) of the project, could influence the result.

In \cite{kohler_automated_2019}, K\"ohler et al. propose a tool for automatic refactoring of OO code to RP which could have been used in order to create a data set where the original code is directly mapped to an RP version. That way, code doing the same thing could have been compared, i.e., the complexity of the problem being solved and the coding style would be the same. At the time of this investigation, the Eclipse plugin called 2rx\footnote{https://github.com/stg-tud/rxrefactoring-code} only refactors a certain set of asynchronous constructs such as \textit{Future} and \textit{SwingWorker}. The generated code wraps the original code into an instance of \textit{RxObserver} extended with the interface of the construct it wraps. As a consequence, the refactored version will always be more complex and is unlikely to yield a fair comparison between OO and RP from e.g. a readability stand point.
The circumstances outlined above led us to the decision to perform a case study translating an existing open source project giving us two versions of the same program. It must be pointed out that this method too has potential flaws which we discuss in Section \ref{threats}, Threats to Validity.

\subsection{Data Analysis}
The refactored methods were analysed using static analysis tools. CYC was measured using PMD\footnote{https://pmd.github.io/}, a popular open source tool, while readability was assessed utilizing the tools provided by the authors\footnote{\url{http://www.arrestedcomputing.com/readability/}, and \url{https://dibt.unimol.it/report/readability/}, respectively.}. All measurements were done per method and mean values over all refactored methods for each version of the system were calculated in order to compare both versions. The values calculated for readability were correlated with the five features of highest predictive power in the B\&W model. Similarly, the values calculated for CYC were correlated with the number of control flow statements.

\section{Results}
\label{results}
In this section, we present a summary of all the data derived from the static analysis. The measurements were performed on a per method basis which resulted in a total of 42 analyzed methods in 5 classes. The diagrams below illustrate a comparison between the refactored and the original versions. The staples in the diagrams represent means over all methods for the individual measures while the black lines describe the standard deviation. Furthermore, the five most influential features of B\&W metric model and the control flow statements which CYC uses are extracted from each method and presented below. 

\subsection{Readability Score}
\begin{figure}[ht]
  \centering
  \includegraphics[width=\linewidth]{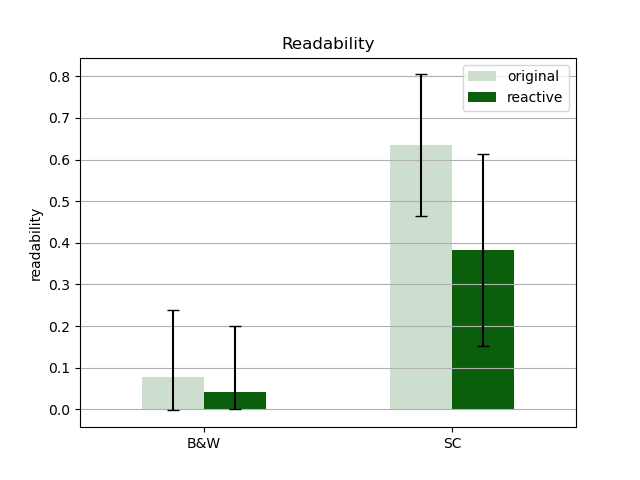}
  \caption{Readability scores given as a mean value for all refactored methods calculated by the B\&W metric and the SC metric on the original code and the refactored version respectively. SC predicts a higher readability overall and a lower decrease in readability for the RP version compared to the original.}
  \label{fig:readabilities}
\end{figure}
In Figure \ref{fig:readabilities}, readability scores are given between 0 (worst readability) and 1 (perfect readability). The SC metric suggest a reduction of the average readability of the refactored methods by 39\%, whilst the B\&W metric suggests a 47\% reduction. Although both metrics agree on reduced readability in the RP version, the SC metric indicates overall higher readability for both OOP and RP than the B\&W metric. 

\subsection{Cyclomatic Complexity}
\begin{figure}[ht]
  \centering
  \includegraphics[width=\linewidth]{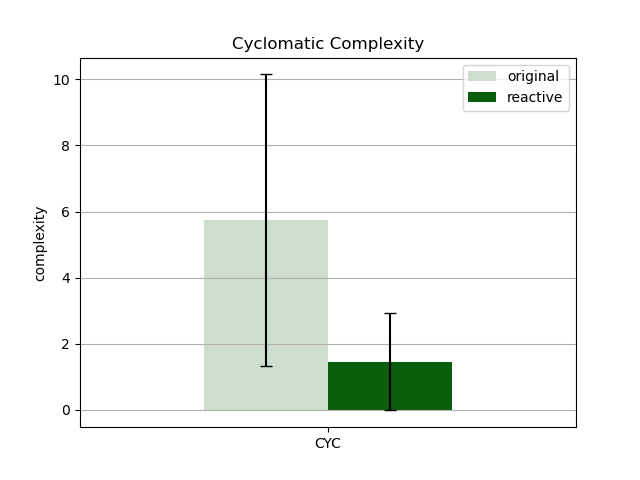}
  \caption{Average cyclomatic complexity value for all refactored methods on the original code and the refactored version respectively. Refactoring to a reactive style has decreased complexity significantly.}
  \label{fig:cyclomatic_complexity}
\end{figure}

\begin{figure}[ht]
  \centering
  \includegraphics[width=\linewidth]{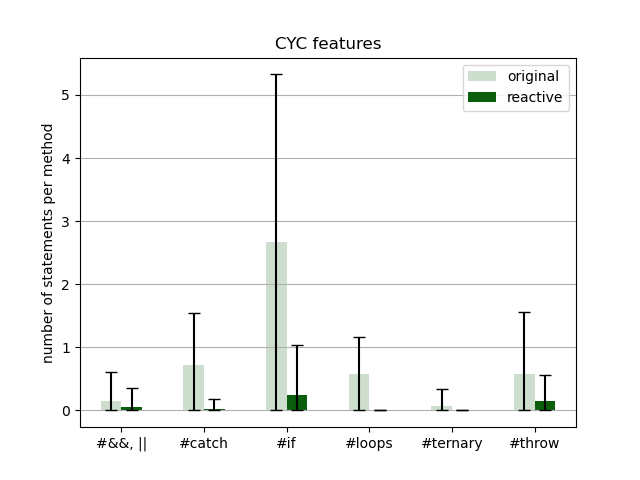}
  \caption{Average numbers of occurrences per method of the features used to calculate cyclomatic complexity in the original code and the refactored version respectively. The features are logical and\slash or operators, catch blocks, if-statements, iterations (loops), ternary operators and throw statements.}
  \label{fig:cyc_features}
\end{figure}

The complexity of each version was calculated using CYC. As can be seen in Figure \ref{fig:cyclomatic_complexity}, an average reduction of $75\%$ in method-level complexity was measured on the refactored version compared to the original. The average numbers of loops (\textit{for}, \textit{while}), if-statements, catch-blocks, binary logical operators, ternary operators and throw-statements per method were counted from the project files (see Figure \ref{fig:cyc_features}). These are all control flow statements used to calculate CYC.

\subsection{Features}
\begin{figure}[ht]
  \centering
  \includegraphics[width=\linewidth]{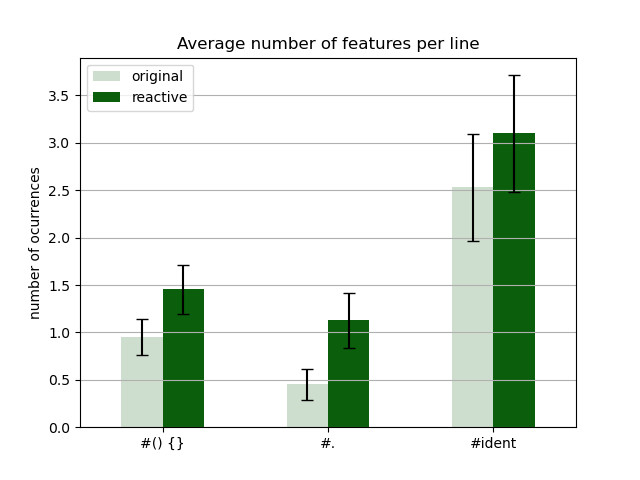}
  \caption{Average number of identifiers, stops and brackets/parentheses per line in the original code and the refactored version. The refactored code contains, on average, more identifiers due to the addition of RxJava operators. The higher number of stops and parentheses is caused by wrapping original code in chained RxJava operators, each such operator adding one stop and and a pair of parentheses}
  \label{fig:avg_features}
\end{figure}

The extracted feature readings of greatest predictive power in the B\&W model are summarized in Table \ref{tab:features} and discussed here. The numbers of stops, parentheses and brackets were counted per method and averaged per line. The refactored version shows an increase by 150\% of stops and an increase of 53\% of parentheses and brackets compared to the original version (see Figure \ref{fig:avg_features}).
In Java, identifiers are names of variables, methods, classes, packages and interfaces that are contained within a class. The average number of identifiers per line were calculated for each method. Figure \ref{fig:avg_features} shows that the number of identifiers increases in average by 22\% compared to the original version.

\begin{table}[ht]
\centering
\begin{tabular}{l|r|r|r}
    \textbf{Measure} & \textbf{OO} ($\mu \pm \sigma$) & \textbf{RP} ($\mu \pm \sigma$) & diff. (\%) \\
    \hline
    \hline
     CYC &  $5.74 \pm 4.41$ & $1.45 \pm 1.47$ & $-75$ \\
     B\&W score & 0.08 $\pm$ 0.16 & 0.04 $\pm$ 0.16 & $-47$ \\ 
     SC score & 0.63 $\pm$ 0.17 & 0.38 $\pm$ 0.23 & $-40$ \\
     number of ids. & 2.53 $\pm$ 0.56 & 3.10 $\pm$ 0.61 & $+22$ \\
     avg line length & 30.56 $\pm$ 6.10 & 34.44 $\pm$ 6.11 & $+13$ \\
     max line length & 76.43 $\pm$ 18.74 & 80.55 $\pm$ 23.98 & $+5.39$ \\
     br. and par. count & 0.95 $\pm$ 0.19 & 1.45 $\pm$ 0.26 & $+53$ \\
     stop count & 0.45 $\pm$ 0.16 & 1.13 $\pm$ 0.29 & $+150 $ \\
     \hline
     \hline
\end{tabular}
\caption{A summary of the measured quantities for code complexity and readability for both versions, object oriented and reactive including the difference in percent.}
\label{tab:features}
\end{table}

\begin{figure}[ht]
  \centering
  \includegraphics[width=\linewidth]{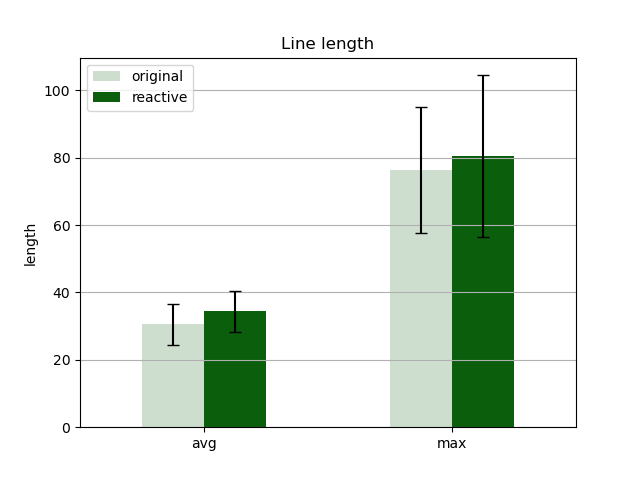}
  \caption{Max line length and average line length in the original code and the refactored version, respectively. No significant difference has been detected.}
  \label{fig:line_length}
\end{figure}

The longest line and the average length per line were calculated with barely any notable difference (see Figure \ref{fig:line_length}).

% \begin{figure}[ht]
%   \centering
%   \includegraphics[width=\linewidth]{sc_most_readable.png}
%   \caption{The ten least negatively affected methods in their reactive version, as measured using the SC metric, and the changes in the five B\&W key features compared to the original version. The readability of the two leftmost methods have actually increased.}
%   \label{fig:sc_most_readable}
% \end{figure}

% \begin{figure}[ht]
%   \centering
%   \includegraphics[width=\linewidth]{sc_least_readable.png}
%   \caption{The ten most negatively affected methods in their reactive version, as measured using the SC metric, and the changes in the five B\&W key features compared to the original version.}
%   \label{fig:sc_least_readable}
% \end{figure}

% Considering SC and the ten methods that were the least negatively affected, as well as the ten methods that were the most negatively affected by the refactoring, we here show the features deviations from the original. In Figures \ref{fig:sc_most_readable} and \ref{fig:sc_least_readable} each method is labeled with a percentage representing the change in readability with SC compared to the original version. The colored dots are the B\&W feature deviations. In Figure \ref{fig:sc_most_readable}, the two leftmost methods have seen an increase in SC readability in the refactored version, while the eight rightmost methods have seen a decrease.

\section{Discussion}
\label{discussion}

In this section, we examine the results to address our research questions. Furthermore, we consider these answers when relating our findings to existing studies. Finally, we discuss related threats to validity.

\emph{RQ1. Using state-of-the-art readability metrics, to what degree can we measure a difference in readability between OOP and RP?}

To answer this question, we considered the B\&W and SC metrics. We saw a big difference in the final scores between the two versions, where RP was shown to yield a lower readability score using both metrics. The SC metric indicates a smaller difference in readability for the OOP version than the B\&W metric. Furthermore, SC yields a significantly higher readability score on both RP and OOP.

\emph{RQ2. What features, if any, in current code readability metrics must be modified or excluded in order to make them applicable to RP?}

The B\&W metric consists of 25 features that the tool considers when evaluating source code. Two out of the five features of highest predictive power in the model are heavily used in RP where a stream consists of a chain of method calls. This naturally gives a disadvantage to RP compared to the OO version with respect to these features, and in extension to the B\&W metric. For instance, regular for and while loops are features that have only a small effect on the score, whereas a corresponding loop in RP such as the  \textit{takeWhile()} operator is an identifier and contains both a stop and a pair of parentheses and therefore has a much stronger negative influence on the score. Consider, for example, the sendToTCP method from the original code in Figure \ref{fig:sendToTcp_orig}. Readability scores are significantly lower for the refactored version (see \ref{fig:sendToTcp_rp}), while, at least to us, it looks more readable (a claim that would have to be substantiated in future studies with humans).

\begin{figure*}[ht]
    \centering
    \begin{subfigure}[t]{\columnwidth}
        \centering
        \begin{lstlisting}[language=Java,
                    basicstyle=\ttfamily\scriptsize,
                    mathescape=true,
                    xrightmargin=2mm]
public void sendToTCP (int connectionID, Object object) 
{
	Connection[] connections = this.connections;
	for (int i = 0, n = connections.length; i < n; i++) {
		Connection connection = connections[i];
		if (connection.id == connectionID) {
			connection.sendTCP(object);
			break;
		}
	}
}
        \end{lstlisting}
    \caption{Original (object oriented)}
    \label{fig:sendToTcp_orig}      
    \end{subfigure}
    \hfill
    \begin{subfigure}[t]{\columnwidth}
        \centering
        \begin{lstlisting}[language=Java,
                            firstnumber=1,
                            mathescape=true,
                            basicstyle=\ttfamily\scriptsize,
                            xrightmargin=2mm]]
public void sendToTCP (int connectionID, Object object) 
{
	Observable.fromArray(this.connections)
			.filter(c -> c.id == connectionID)
			.firstElement()
			.subscribe(c -> c.sendTCP(object));
}
        \end{lstlisting}
        \vspace{1.06cm}
        \caption{Refactored (reactive)}
        \label{fig:sendToTcp_rp}      
    \end{subfigure}
    \caption{The original \textit{sendToTCP} method (left) compared to the refactored version (right) using the reactive code-style. The refactored version has a much lower readability (0.054\% as measured by B\&W and 53.00\% as measured by SC) score than the original one (2.89\% as measured by B\&W and 77.39\% as measured by SC). The reduction of code-lines does not carry a positive effect on readability. However, extra identifiers, stops, brackets and parentheses have a strong negative impact on readability in our findings.}
    \label{fig:sendToTcp}
\end{figure*}

% \begin{figure*}[ht]
%     \centering
%     \begin{subfigure}[t]{\columnwidth}
%         \centering
%         \includegraphics[width=8.1cm]{sendToTCP_orig.PNG}
%         \caption{Original (object oriented)}
%         \label{fig:sendToTcp_orig}      
%     \end{subfigure}
%     \begin{subfigure}[t]{\columnwidth}
%         \centering
%         \vspace{-3.65cm}
%         \includegraphics[width=8.1cm]{sendToTCP_rp.PNG}
%         \vspace{1.35cm}
%         \caption{Refactored (reactive)}
%         \label{fig:sendToTcp_rp}      
%     \end{subfigure}
%     \caption{The original \textit{sendToTCP} method (left) compared to the refactored version (right) using the reactive code-style. The refactored version has a much lower readability (0.054\% as measured by B\&W and 53.00\% as measured by SC) score than the original one (2.89\% as measured by B\&W and 77.39\% as measured by SC). The reduction of code-lines does not carry a positive effect on readability. However, extra stops, brackets and parentheses have a strong negative impact on readability in our findings.}
%     \label{fig:sendToTcp}
% \end{figure*}

The reduction in the number of lines has little impact on the readability in the metrics used, while the extra identifiers (RxJava operators), stops, brackets and parentheses have a strong impact. Both metrics count these characters as average occurrences per line. This means, for example, that the original methods have an average of 0.4 stops (periods) per line while the RP version has an average of 1.2 stops per line. This is an increase of 200\%. We argue that this does not reflect the difference in readability and that the conciseness and the descriptive (declarative) nature of the reactive version is not taken into account. Consequentially, we consider the B\&W metrics inapplicable for RP.

The SC metric was developed by combining existing known readability metrics, i.e. the ones presented by B\&W, Posnett, Dorn and Scalabrino’s first model (using only textual features). It was found by the authors to achieve higher predictive accuracy for readability with its combination of structural and textual features. Considering the work of Salvaneschi et al., where RP was found to be more comprehensible than OOP, partly due to better readability, our results confirm that SC performs better than B\&W since its readability deviations are less pronounced. However, the structural features of the B\&W metric are an integral part of SC, which renders even SC less applicable for RP than OOP. Regarding the slightly better results in favour of RP readability with SC, we have been unable to trace the specific reasons for this and can only speculate that it has to do with the spatial and visual features of the Dorn model that may favour the declarative style of RP. 

\emph{RQ3. How does RP, with its declarative nature, affect code complexity compared  to OOP?}

Examining the score of CYC, RP has shown to decrease the code complexity significantly compared to the original version. CYC calculates its score simply by counting the number of potential paths through the code where the number of flow statements determine the final score. Studying the flow statement diagram from the previous section, we can detect important differences in the two versions. The RP version has replaced the loops with method calls (operators) from the RxJava API and by that minimized the use of loops. The number of if-statements has been reduced to some extent by filters. In addition, utilizing RxJava's error channel, several try/catch-blocks have been removed. The above-mentioned are all considered as flow statements which affect the final complexity score.

\subsection{Contrasting Results with Related Work}
Our study can be contrasted with the findings from the empirical study in \cite{salvaneschi_positive_2017}  where the authors claimed that RP increases source code comprehension when directly compared to OOP. Moreover, they pointed out that their human subjects found RP to be more comprehensive due to increased readability. The findings from our case study suggest the contrary but do raise some doubts. By answering the second research question, we can conclude that the current metrics used to measure code readability are not applicable to RP, thus making a fair comparison between OOP and RP hard to impossible. Furthermore, the study conducted by Pantiuchina et al. shows that the B\&W metric and Scalabrino’s readability model are unable to recognize any code improvements, which questions the legitimacy of the metrics.

To the best of our knowledge, no studies comparing code complexity between OOP and RP have been conducted. CYC can be calculated on each version and still make up for a fair comparison between the projects. It acts independently and does not value any unique features of certain paradigms differently. Referring back to the previous section, we conclude that RP seems to lower source code complexity.

\subsection{Validity Threats}
\label{threats}
It has to be acknowledged that the refactored RP version of the code used in this study may not be written in a way that makes it representative of all reactive programming nor can it be claimed to utilize the full potential of the RxJava API. As a case study the work cannot be expected to cover the entire state of reactive programming, but serve as an indicator and be used as a motivation for future research. 

The manual refactoring conducted by the authors could be considered susceptible to research bias, i.e. that special efforts could have been made to favour readability. This threat was mitigated by, firstly, at the time of the refactoring not having any detailed information about which features were actually used in the two readability models, secondly, by refactoring each method as a 'black box', meaning only its inner structure has been changed and no functionality has been neither added nor removed. For example, no unwieldy logging statements were extracted from the methods and placed elsewhere, which would clearly have improved the methods readability score, but not provided a fair comparison between the two versions. 

% The refactoring strategy used, where each statement is mapped to its RxJava counterpart, means that there is basically only one way to refactor a code segment, e.g. looping through a Collection. 
\section{Conclusions}
\label{conclusions}
The purpose of this study was to gain a better understanding of whether current complexity and readability metrics are suitable for non-traditional coding-styles and paradigms that are growing in use. We therefore compared a popular complexity metric \cite{mccabe_complexity_1976} and two state-of-the-art readability metrics from \cite{buse_learning_2010, scalabrino_comprehensive_2018} for an open source object-oriented project and a fork that we refactored using a popular library for reactive programming as an example of that category. Notable differences in both complexity and readability were found. We found that the cyclomatic complexity in the reactive version is significantly lower, which coincides with the ambitions around reactive programming. The metric was suggested over 40 years ago but may extend well to the reactive paradigm. However, the two readability metrics from the literature indicate a significant reduction in readability which may seem counter-intuitive, and which we challenged, also by the example given above, as bad assessment. A potential source of error here is the large emphasis on structural source-code features in the existing readability metrics which may render them unsuitable for reactive code with it's chaining character and use of lambdas. The B\&W metric from \cite{buse_learning_2010} solely relies on structural features while the one from \cite{scalabrino_comprehensive_2018} combines structural and textual features resulting in a still notable but less severe negative readability impact in reactive programming. In essence, none of the metrics captures readability in a desired way.

In order to better understand the impact of programming paradigms on code readability, broader studies on both reactive but also other functional or declarative coding styles are required. Multiple-dispatch should be mentioned here too as a relevant object of investigation that gains traction, for instance through the Julia programming language. A more in-depth analysis of the existing metrics and their characteristics may further lead to their improvement for code segments of various styles. In addition, languages such as Java and C\# support the arbitrary combination of styles within the same project which may require work on detecting and differentiating code-styles within a single project. Meaningful and objectively correct readability characteristics, through e.g. linting, could prove to be invaluable in a development environment. How to address the question of assessing source-code readability objectively is a so far unanswered question, which is why we see much research in that realm being called for.

% Note that the IEEE does not put floats in the very first column
% - or typically anywhere on the first page for that matter. Also,
% in-text middle ("here") positioning is typically not used, but it
% is allowed and encouraged for Computer Society conferences (but
% not Computer Society journals). Most IEEE journals/conferences use
% top floats exclusively. 
% Note that, LaTeX2e, unlike IEEE journals/conferences, places
% footnotes above bottom floats. This can be corrected via the
% \fnbelowfloat command of the stfloats package.

% trigger a \newpage just before the given reference
% number - used to balance the columns on the last page
% adjust value as needed - may need to be readjusted if
% the document is modified later
%\IEEEtriggeratref{8}
% The "triggered" command can be changed if desired:
%\IEEEtriggercmd{\enlargethispage{-5in}}

% references section

% can use a bibliography generated by BibTeX as a .bbl file
% BibTeX documentation can be easily obtained at:
% http://mirror.ctan.org/biblio/bibtex/contrib/doc/
% The IEEEtran BibTeX style support page is at:
% http://www.michaelshell.org/tex/ieeetran/bibtex/
\balance
\bibliographystyle{IEEEtran}
% argument is your BibTeX string definitions and bibliography database(s)
\bibliography{reactive_quality}

% Generated by IEEEtran.bst, version: 1.12 (2007/01/11)
\begin{thebibliography}{10}
\providecommand{\url}[1]{#1}
\csname url@samestyle\endcsname
\providecommand{\newblock}{\relax}
\providecommand{\bibinfo}[2]{#2}
\providecommand{\BIBentrySTDinterwordspacing}{\spaceskip=0pt\relax}
\providecommand{\BIBentryALTinterwordstretchfactor}{4}
\providecommand{\BIBentryALTinterwordspacing}{\spaceskip=\fontdimen2\font plus
\BIBentryALTinterwordstretchfactor\fontdimen3\font minus
  \fontdimen4\font\relax}
\providecommand{\BIBforeignlanguage}[2]{{%
\expandafter\ifx\csname l@#1\endcsname\relax
\typeout{** WARNING: IEEEtran.bst: No hyphenation pattern has been}%
\typeout{** loaded for the language `#1'. Using the pattern for}%
\typeout{** the default language instead.}%
\else
\language=\csname l@#1\endcsname
\fi
#2}}
\providecommand{\BIBdecl}{\relax}
\BIBdecl

\bibitem{buse_learning_2010}
R.~P. Buse and W.~R. Weimer, ``Learning a {Metric} for {Code} {Readability},''
  \emph{IEEE Transactions on Software Engineering}, vol.~36, no.~4, pp.
  546--558, Jul. 2010, conference Name: IEEE Transactions on Software
  Engineering.

\bibitem{posnett_simpler_2011}
D.~Posnett, A.~Hindle, and P.~Devanbu, ``A simpler model of software
  readability,'' in \emph{Proceedings of the 8th {Working} {Conference} on
  {Mining} {Software} {Repositories}}, ser. {MSR} '11.\hskip 1em plus 0.5em
  minus 0.4em\relax Waikiki, Honolulu, HI, USA: Association for Computing
  Machinery, May 2011, pp. 73--82.

\bibitem{alawad_empirical_2019}
\BIBentryALTinterwordspacing
D.~Alawad, M.~Panta, M.~Zibran, and M.~R. Islam, ``An {Empirical} {Study} of
  the {Relationships} between {Code} {Readability} and {Software}
  {Complexity},'' \emph{arXiv:1909.01760 [cs]}, vol. abs/1909.01760, Aug. 2019,
  arXiv: 1909.01760. [Online]. Available: \url{http://arxiv.org/abs/1909.01760}
\BIBentrySTDinterwordspacing

\bibitem{noauthor_iso_nodate}
\BIBentryALTinterwordspacing
{International Organization for Standardization}, ``{ISO} 25010,'' 2017.
  [Online]. Available:
  \url{https://iso25000.com/index.php/en/iso-25000-standards/iso-25010}
\BIBentrySTDinterwordspacing

\bibitem{chidamber_metrics_1994}
S.~Chidamber and C.~Kemerer, ``A metrics suite for object oriented design,''
  \emph{IEEE Transactions on Software Engineering}, vol.~20, no.~6, pp.
  476--493, Jun. 1994, conference Name: IEEE Transactions on Software
  Engineering.

\bibitem{redol_object-oriented_nodate}
F.~B. Abreu and R.~Carapu{\c{c}}a, ``Object-oriented software engineering:
  Measuring and controlling the development process,'' in \emph{Proceedings of
  the 4th international conference on software quality}, vol. 186, MLean, VA,
  USA, 1994.

\bibitem{salvaneschi_positive_2017}
G.~Salvaneschi, S.~Proksch, S.~Amann, S.~Nadi, and M.~Mezini, ``On the
  {Positive} {Effect} of {Reactive} {Programming} on {Software}
  {Comprehension}: {An} {Empirical} {Study},'' \emph{IEEE Transactions on
  Software Engineering}, vol.~43, no.~12, pp. 1125--1143, Dec. 2017, conference
  Name: IEEE Transactions on Software Engineering.

\bibitem{dorn_general_2012}
J.~Dorn, ``A general software readability model,'' \emph{MCS Thesis available
  from (http://www. cs. virginia. edu/weimer/students/dorn-mcs-paper. pdf)},
  vol.~5, pp. 11--14, 2012.

\bibitem{scalabrino_improving_2016}
S.~Scalabrino, M.~Linares-Vasquez, D.~Poshyvanyk, and R.~Oliveto,
  ``\BIBforeignlanguage{en}{Improving code readability models with textual
  features},'' in \emph{\BIBforeignlanguage{en}{2016 {IEEE} 24th
  {International} {Conference} on {Program} {Comprehension} ({ICPC})}}.\hskip
  1em plus 0.5em minus 0.4em\relax Austin, TX, USA: IEEE, May 2016, pp. 1--10.

\bibitem{scalabrino_comprehensive_2018}
S.~Scalabrino, M.~Linares‐Vásquez, R.~Oliveto, and D.~Poshyvanyk,
  ``\BIBforeignlanguage{en}{A comprehensive model for code readability},''
  \emph{\BIBforeignlanguage{en}{Journal of Software: Evolution and Process}},
  vol.~30, no.~6, 2018.

\bibitem{mduniversity_code_2017}
{M.D.University}, A.~Pahal, and R.~S.~Chillar, ``\BIBforeignlanguage{en}{Code
  {Readability}: {A} {Review} of {Metrics} for {Software} {Quality}},''
  \emph{\BIBforeignlanguage{en}{International Journal of Computer Trends and
  Technology}}, vol.~46, no.~1, pp. 1--4, Apr. 2017.

\bibitem{bainomugisha_survey_2013}
E.~Bainomugisha, A.~L. Carreton, T.~v. Cutsem, S.~Mostinckx, and W.~d. Meuter,
  ``A survey on reactive programming,'' Aug. 2013.

\bibitem{banken_debugging_2018}
H.~Banken, E.~Meijer, and G.~Gousios, ``Debugging {Data} {Flows} in {Reactive}
  {Programs},'' in \emph{2018 {IEEE}/{ACM} 40th {International} {Conference} on
  {Software} {Engineering} ({ICSE})}.\hskip 1em plus 0.5em minus 0.4em\relax
  New York, NY, USA: Association for Computing Machinery, May 2018, pp.
  752--763.

\bibitem{tashtoush_impact_2013}
Y.~Tashtoush, Z.~Odat, I.~Alsmadi, and M.~Yatim, ``Impact of {Programming}
  {Features} on {Code} {Readability},'' \emph{Computer Science Faculty
  Publications}, vol.~7, no.~6, Jan. 2013.

\bibitem{flesch_new_1948}
R.~Flesch, ``A new readability yardstick,'' \emph{Journal of Applied
  Psychology}, vol.~32, no.~3, pp. 221--233, 1948, place: US Publisher:
  American Psychological Association.

\bibitem{mccabe_complexity_1976}
T.~McCabe, ``A {Complexity} {Measure},'' \emph{IEEE Transactions on Software
  Engineering}, vol. SE-2, no.~4, pp. 308--320, Dec. 1976, conference Name:
  IEEE Transactions on Software Engineering.

\bibitem{halstead_elements_1977}
M.~H. Halstead \emph{et~al.}, \emph{Elements of software science}.\hskip 1em
  plus 0.5em minus 0.4em\relax Elsevier New York, 1977, vol.~7.

\bibitem{shepperd_critique_1988}
\BIBentryALTinterwordspacing
M.~Shepperd, ``\BIBforeignlanguage{en}{A critique of cyclomatic complexity as a
  software metric},'' \emph{\BIBforeignlanguage{en}{Software Engineering
  Journal}}, vol.~3, no.~2, pp. 30--36, Mar. 1988, publisher: IET Digital
  Library. [Online]. Available:
  \url{https://digital-library.theiet.org/content/journals/10.1049/sej.1988.0003}
\BIBentrySTDinterwordspacing

\bibitem{pizka_how_2007}
M.~Pizka and F.~Deissenb{\"o}ck, ``How to effectively define and measure
  maintainability,'' \emph{SMEF 2007}, p.~93, 2007.

\bibitem{noauthor_cyclomatic_nodate}
M.~M.~S. Sarwar, S.~Shahzad, and I.~Ahmad, ``Cyclomatic complexity: The nesting
  problem,'' in \emph{Eighth International Conference on Digital Information
  Management (ICDIM 2013)}.\hskip 1em plus 0.5em minus 0.4em\relax IEEE, 2013,
  pp. 274--279.

\bibitem{pantiuchina_improving_2018}
J.~Pantiuchina, M.~Lanza, and G.~Bavota, ``Improving {Code}: {The} ({Mis})
  {Perception} of {Quality} {Metrics},'' in \emph{2018 {IEEE} {International}
  {Conference} on {Software} {Maintenance} and {Evolution} ({ICSME})}, Sep.
  2018, pp. 80--91.

\bibitem{fakhoury_improving_2019}
\BIBentryALTinterwordspacing
S.~Fakhoury, D.~Roy, A.~Hassan, and V.~Arnaoudova,
  ``\BIBforeignlanguage{en}{Improving {Source} {Code} {Readability}: {Theory}
  and {Practice}},'' in \emph{\BIBforeignlanguage{en}{2019 {IEEE}/{ACM} 27th
  {International} {Conference} on {Program} {Comprehension} ({ICPC})}}.\hskip
  1em plus 0.5em minus 0.4em\relax Montreal, QC, Canada: IEEE, May 2019, pp.
  2--12. [Online]. Available:
  \url{https://ieeexplore.ieee.org/document/8813254/}
\BIBentrySTDinterwordspacing

\bibitem{scalabrino_automatically_2019}
S.~Scalabrino, G.~Bavota, C.~Vendome, M.~Linares-Vásquez, D.~Poshyvanyk, and
  R.~Oliveto, ``Automatically assessing code understandability,'' \emph{{IEEE}
  Transactions on Software Engineering}, pp. 1--1, 2019, conference Name:
  {IEEE} Transactions on Software Engineering.

\bibitem{fakhoury_effect_2018}
\BIBentryALTinterwordspacing
S.~Fakhoury, Y.~Ma, V.~Arnaoudova, and O.~Adesope, ``The effect of poor source
  code lexicon and readability on developers' cognitive load,'' in
  \emph{Proceedings of the 26th Conference on Program Comprehension}, ser. ICPC
  '18.\hskip 1em plus 0.5em minus 0.4em\relax New York, NY, USA: Association
  for Computing Machinery, 2018, p. 286–296. [Online]. Available:
  \url{https://doi-org.proxybib.miun.se/10.1145/3196321.3196347}
\BIBentrySTDinterwordspacing

\bibitem{mannan_towards_2018}
\BIBentryALTinterwordspacing
U.~A. Mannan, I.~Ahmed, and A.~Sarma, ``Towards understanding code readability
  and its impact on design quality,'' in \emph{Proceedings of the 4th {ACM}
  {SIGSOFT} International Workshop on {NLP} for Software Engineering}, ser.
  {NL}4SE 2018.\hskip 1em plus 0.5em minus 0.4em\relax Association for
  Computing Machinery, 2018, pp. 18--21. [Online]. Available:
  \url{http://doi.org/10.1145/3283812.3283820}
\BIBentrySTDinterwordspacing

\bibitem{mendoncca2020understanding}
W.~L.~M. Mendon{\c{c}}a, R.~B. de~Almeida, J.~Fortes, F.~V. Lopes,
  D.~Marc{\'\i}lio, E.~D. Canedo, F.~Lima, and J.~Saraiva, ``Understanding the
  impact of introducing lambda expressions in java programs,'' \emph{Journal of
  Software Engineering Research and Development}, vol.~8, pp. 8--1, 2020.

\bibitem{vanags_perfect_2018}
M.~Vanags and R.~Cevere, ``\BIBforeignlanguage{en}{The {Perfect} {Lambda}
  {Syntax}},'' \emph{\BIBforeignlanguage{en}{Baltic Journal of Modern
  Computing}}, vol.~6, no.~1, 2018.

\bibitem{martin_clean_2008}
R.~C. Martin, \emph{Clean {Code}: {A} {Handbook} of {Agile} {Software}
  {Craftsmanship}}, 1st~ed.\hskip 1em plus 0.5em minus 0.4em\relax USA:
  Prentice Hall PTR, 2008.

\bibitem{kohler_automated_2019}
M.~Köhler and G.~Salvaneschi, ``Automated {Refactoring} to {Reactive}
  {Programming},'' in \emph{2019 34th {IEEE}/{ACM} {International} {Conference}
  on {Automated} {Software} {Engineering} ({ASE})}, Nov. 2019, pp. 835--846.

\end{thebibliography}

% that's all folks
\end{document}